\documentclass[runningheads]{llncs}
\usepackage{svg}
\usepackage{epsfig}

\usepackage{adjustbox}
\usepackage{amsmath,amssymb,amsfonts}
\usepackage{graphicx}
\usepackage{textcomp}\usepackage{amssymb}
\usepackage{pifont}
\usepackage{algorithm}
\usepackage{algcompatible}
\usepackage{lipsum}
\usepackage{algorithm}

\usepackage{makecell}

\newcommand{\xmark}{\ding{55}}
\usepackage{xcolor}
\usepackage{multicol}
\usepackage{multirow}
\usepackage{subcaption}
\usepackage{enumitem}
\usepackage{balance}
\usepackage{booktabs}
\usepackage{array}
\newcolumntype{x}[1]{>{\centering\arraybackslash\hspace{0pt}}p{#1}}

\usepackage{pst-all}
\usepackage{pstricks-add}
\usepackage{float}

\usepackage{pgfplots}

\usepackage{tikz}
\usepackage{tikzscale}
\usepackage{transparent}
\usetikzlibrary{snakes}
\pgfplotsset{
  /pgfplots/line legend with two nodes/.style 2 args={
    legend image code/.code={
      \draw[##1,no markers]
        plot coordinates {
        (0cm,0cm)
        (0.3cm,0cm)
        (0.6cm,0cm)
      }
      node[pos=0,#1]{}
      node[#2]{};%
    }
  }
}

\usepackage{algorithm}
\usepackage[noend]{algpseudocode}

\begin{document}

\title{Efficient Network Representation for GNN-based Intrusion Detection}


\author{Hamdi FRIJI\inst{1,2}\and
Alexis OLIVEREAU\inst{1} \and
Mireille SARKISS\inst{2}}
\authorrunning{H. Friji et al.}
%
\institute{CEA, LIST, Communicating Systems Laboratory, F-91191 Gif-sur-Yvette, France
\email{\{hamdi.friji,alexis.olivereau\}@cea.fr} \and
SAMOVAR, Télécom SudParis, Institut Polytechnique de Paris, 91120 Palaiseau, France
\email{\{hamdi\_friji,mireille.sarkiss\}@telecom-sudparis.eu}
}
\maketitle              

\begin{abstract}

\textcolor{black}{The last decades have seen a growth in the number of cyber-attacks with severe economic and privacy damages, which reveals the need for network intrusion detection approaches to assist in preventing cyber-attacks and reducing their risks. 
In this work, we propose a novel network representation as a graph of flows that aims to provide relevant topological information for the intrusion detection task, such as malicious behavior patterns, the relation between phases of multi-step attacks, and the relation between spoofed and pre-spoofed attackers' activities.
In addition, we present a Graph Neural Network (GNN) based-framework responsible for exploiting the proposed graph structure to classify communication flows by assigning them a maliciousness score. The framework comprises three main steps that aim to embed nodes' features and learn relevant attack patterns from the network representation.
Finally, we highlight a potential data leakage issue with classical evaluation procedures and suggest a solution to ensure a reliable validation of intrusion detection systems' performance. 
We implement the proposed framework and prove that exploiting the flow-based graph structure outperforms the classical machine learning-based and the previous GNN-based solutions.}

\end{abstract}


\section{Introduction}

During the last decades, with the emergence of Internet of Things (IoT), cloud, and edge computing, cyber-attacks have increased exponentially in frequency and complexity. Accordingly, adopting the industry 4.0 technologies has opened the doors for attackers to take advantage of many security breaches \cite{9612213}. 
Indeed, some characteristics of industry 4.0 are considered highly appealing targets for cyber-attackers, for example: 1) Industry 4.0 technologies are usually implemented over an old isolated system, thus increasing the attack surface and giving more opportunities to the attackers, 2) a massive number of connected devices are considered potential attack risks, 3) lack of visibility across isolated environments and separate systems can lead to critical security issues.

Cyber-criminals attempt to use these vulnerabilities to carry out malicious activities~\cite{9663537} and gain access to unauthorized information, such as ransomware and man-in-the-middle attacks, disrupt services, such as Distributed Denial of Service (DDoS), Denial of Service (DoS), and botnet attacks, or destructive attacks using malware. The attacks can be costly for companies and governments. For instance, the ransomware attack that scoped JBS S.A company on May 30, 2021, was solved only by paying 11 million dollars to the attackers to regain access to all internal data. The Harris Federation, a school group in the United Kingdom, was hit by ransomware and was threatened to release sensitive data. Therefore, governments and companies support the development of efficient solutions to alleviate the risks of these attacks \cite{9663537}. In addition, cyber-attacks are getting increasingly sophisticated, and hackers employ several techniques to evade the existing Intrusion Detection Systems (IDSs). For this reason, appropriate reactive schemes are needed to go along with the attacks' evolution pace.\\

To ensure more protection and detect these attacks, companies considers two main types of IDSs \cite{Dubey2013ASO}:
\begin{itemize}
    
    \item Signature-based intrusion detection systems: These systems aim to detect attacks by comparing network traffic to predefined patterns of attacks that are already known.

    \item Anomaly-based intrusion detection systems: These systems monitor the traffic to detect any abnormal behavior. They use statistical and machine learning techniques to classify the traffic into normal and anomalous (binary clustering), or into normal and attack(s) (classification). 

\end{itemize}

The existing signature-based IDSs are efficient in detecting the known attacks with a low false-positive rate but fail to detect any new type of attacks~\cite{Dubey2013ASO}. Therefore, during the last few years, researchers started to exploit anomaly detection approaches using Machine Learning (ML) and Deep Learning (DL) algorithms to identify any abnormal behavior in the network.
On the one hand, these techniques show promising results in detecting unseen attacks. Yet, on the other hand, it is difficult to achieve a low false-positive rate using these algorithms. One of the reasons behind this issue is the poor quality of the models being used, partly due to the lack of high-quality datasets. Indeed, these datasets are generally biased towards including mostly regular traffic at the expense of malicious traffic.\\

\textcolor{black}{
This work addresses the problem of detecting cyber-attacks at the network edge using an anomaly-based intrusion detection approach. The detection consists in classifying the communication flows into normal and anomalous. We propose an efficient network representation that provides relevant information about cyber-attacks, enabling Artificial Intelligence (AI) algorithms to detect malicious patterns. Our structure provides important structural information about attackers' behavior, such as the iterative malicious routines, connections between several malicious activities, and the existence of multi-step attacks or distributed attacks.
Moreover, we propose a framework for detecting network-level attacks based on Graph Neural Networks (GNN) to exploit our network representation. The framework comprises three main steps that aim to extract and embed attacks topological features with flow's attributes. The obtained information characterizes the graph's topology with the nodes' attributes and enhances the capabilities of the decision-making module in distinguishing between normal and malicious flows.
The availability of datasets for the flow classification problem and the high complexity of packet-level IDSs endorses our choice to work on a flow-based approach.\\
}

Our main contributions can be summarized as follows:\\
\textcolor{black}{
$\bullet$ We propose a novel flow-based graph structure that provides relevant information about malicious behavior patterns and enables the model to attend a higher accuracy in distinguishing between normal and malicious flows.\\
$\bullet$ We devise a graph-based framework to extract and exploit the spatial information of our network representation to detect malicious flows.\\
$\bullet$ We analyze the currently used validation techniques to ensure a reliable validation of the IDS.\\
$\bullet$ We evaluate our proposed model by comparing it with previous intrusion detection works. We show that our results are promising and outperform the existing ML-based and graph-based works.\\
}

The remainder of this paper is structured as follows. First, the related work is reviewed in Section II. Section III then presents and explains the preliminaries needed to understand the rest of the work. Section IV introduces the proposed framework and its different phases. Section V discusses the existing evaluation techniques. Finally, the obtained results are presented in section VI, and the conclusion is given in section VII.

\section{Related Work}
The problem of intrusion detection has been widely investigated in the literature. We review in this section the most latent works that proposed or used graph structures to perform intrusion detection.

\textcolor{black}{Josep Soler Garrido et al. \cite{9527927} proposed to use relational learning on knowledge graphs to ensure security monitoring and accomplish intrusion detection. They apply ML techniques on knowledge graphs to detect unexpected activity in industrial automation systems. The knowledge graphs provide relevant information, but they cause considerable memory consumption, and due to using IP addresses as an identifier, they can be eluded easily using IP spoofing\footnote{IP address spoofing refers to the creation of Internet Protocol (IP) packets with a false source IP address to evade the detection by intrusion detection systems}.}

\textcolor{black}{The authors of \cite{DBLP:journals/corr/abs-2107-14756} exploited the CICIDS 2017 dataset to create a host-connection graph. They first propose to create a heterogeneous graph where they introduce two types of nodes; the first type represents users, and the second represents the flows. This structure is more complicated to handle, and hence authors were obliged to propose their message-passing \cite{https://doi.org/10.48550/arxiv.2202.11097} procedure in order to be able to exploit their network representation. Moreover, this structure could also be evaded using IP spoofing since it is based on IP addresses as an identifier of users' nodes.}

\textcolor{black}{Yulong Pei et al. \cite{9564233} proposed to use a Graph Convolutional Network (GCN) to address the anomaly detection problem on attributed networks. The same problem is studied in \cite{9634972}, where the authors exploit the node attention mechanism to better obtain the network's embedding representation. Ultimately, they use a multi-layer perceptron algorithm to train data to detect any malicious activity. However, these works use the classical sub-efficient graph structure where nodes represent users and flows are the edges of the graph.
Saber Zerhoudi et al.~\cite{9202751} suggested enhancing intrusion detection systems using zero-shot learning. Their framework aims to improve insider threat detection performance for cases where historical user data is unavailable. Specifically, they used graph embeddings to encode relations from the organization structure.
In \cite{DBLP:journals/corr/abs-2103-16329}, the authors proposed to apply the predefined E-GraphSage algorithm to the previously described classical graph structure to achieve intrusion detection.}

\textcolor{black}{The works in \cite{DBLP:journals/corr/abs-2103-16329,9564233,9634972,9202751} use the classical graph representation which characterizes the users as nodes and communication flows as edges. This structure has several drawbacks, for example, it can be evaded since it is based on IP addresses, a pre-processing phase is required to make it consumable by GNN algorithms, and it does not provide relevant topological information about distributed or multi-steps attacks.}

\textcolor{black}{Our network representation provides more relevant information in comparison with the graph structures of the previously cited works. Moreover, our structure is not affected by IP spoofing and it gives the possibility to link several steps of multi-step attacks.
}

Table~\ref{compare} summarizes the distinguishing features of our network representation compared to the aforementioned papers.

\begin{table}[h]
\caption{\textcolor{black}{High-level comparison of the proposed graph representation features with previous structures used for intrusion detection}}
\centering
\begin{tabular}{>{\centering\arraybackslash}p{3.5cm} 
>{\centering\arraybackslash}p{1.8cm} 
>{\centering\arraybackslash}p{0.6cm} 
>{\centering\arraybackslash}p{0.6cm} 
>{\centering\arraybackslash}p{0.6cm} 
>{\centering\arraybackslash}p{0.6cm}
>{\centering\arraybackslash}p{0.6cm} 
>{\centering\arraybackslash}p{0.6cm} 
>{\centering\arraybackslash}p{0.6cm}}
\toprule
\textbf{Feature}              & \textbf{Proposed structure} & \textbf{~\cite{9527927}} & \textbf{~\cite{DBLP:journals/corr/abs-2103-16329}} & \textbf{~\cite{9564233}}  & \textbf{~\cite{DBLP:journals/corr/abs-2107-14756}}  & \textbf{~\cite{9634972}}  & \textbf{~\cite{9202751}} \\\midrule
Exploitability by GNN-based models      & \checkmark       & \checkmark        & \checkmark                    & \checkmark         & \checkmark         & \checkmark         & \checkmark                     \\ \midrule
Flow-based graph structure      & \checkmark       & \xmark        & \xmark                      & \xmark             & \xmark             & \xmark             & \xmark                      \\ \midrule

Relevant topological information & \checkmark        & \checkmark        & \xmark                    & \xmark          & \checkmark        & \xmark         & \xmark                       \\ \midrule
Cover Multi-step attacks         & \checkmark        & \xmark        & \xmark                     & \xmark             & \xmark             & \xmark             & \xmark                      \\ \midrule
IP Spoofing immunity      & \checkmark       & \xmark        & \xmark                      & \xmark             & \xmark             & \xmark             & \xmark                      \\ \midrule

Simplicity of exploitation         & \checkmark        & \xmark        & \xmark                     & \xmark             & \xmark             & \xmark             & \xmark                      \\ \midrule

Lower memory consumption  & \checkmark        & \xmark        & \checkmark                     & \xmark             & \xmark             & \checkmark             & \checkmark                     \\ 

\bottomrule
    
\end{tabular}
\label{compare}  
\end{table}

\section{Preliminaries}
This section introduces and explains some concepts needed for a better understanding of the paper.
\subsection{NIDS Problem Statement}

\begin{figure}[t!]
    \centerline{\includegraphics[width=8.5cm]{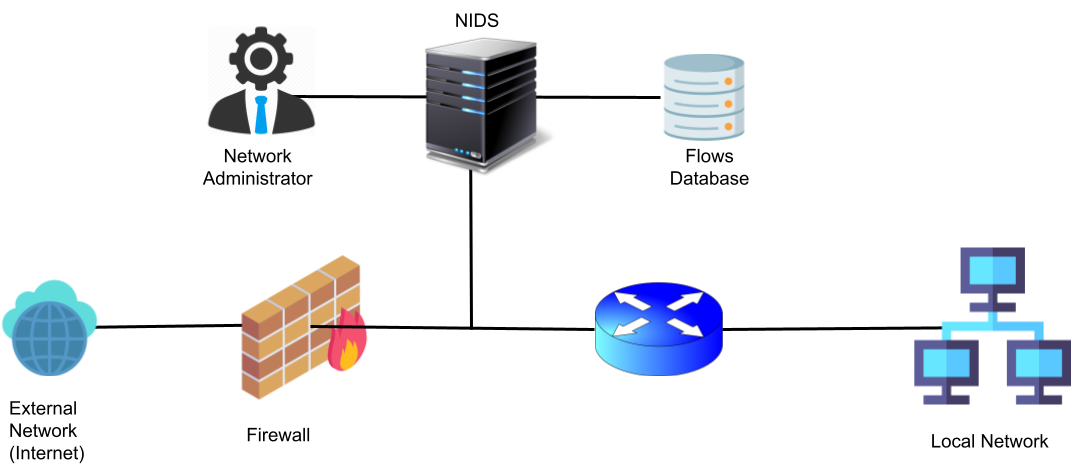}}\vspace{-0.2cm}
    \caption{Illustration of the network intrusion detection}
    \label{fig:network}\vspace{-0.5cm}
\end{figure}

Our objective is to propose a Network-level IDS (NIDS), as illustrated in Fig.~\ref{fig:network}. The NIDS's role is to classify the network flows into normal and malicious and report any abnormal behavior to the system administrator (SA). The SA is then responsible for analyzing the reported network flows and reacting to prevent potential attacks.
Meanwhile, the flows classified as malicious are stored in a specific dataset for further processing.
The NIDS collects packets from the network in the firewall's internal interface. The collected packets will then be transformed into communication flows, identified by the source Internet Protocol (IP) address, source port number, destination IP address, destination port number, and timestamp. 
Several statistical features related to the packets can be extracted during the flows creation, such as the used protocols, SSL activity, and connection activity. These features are assigned to flows as attributes.
\subsection{General Graph Definition}

In this study, we define a graph as a mathematical structure $G=(V,E,R,X)$ where $V= \{v_1,v_2,\ldots,v_N\} $ is the set of nodes in the graph, $E=\{e_1,e_2,\ldots,e_{N^\prime}\}$ is the edge set associated with $V$, and $R=\{r_1,r_2,\ldots,r_{N^{\prime}}\}$ represents the edges' attributes of size $N^{\prime\prime}$. $N=|V|$ and ${N^\prime}=|E|$ denote the total number of nodes and edges, respectively. $X=\{x_1,x_2,\ldots,x_{N^{\prime\prime}}\}$ is the set of node features of size $N^{\prime\prime}$.

\subsection{Graph-Line Representation}
\label{sec:graph-line}

The graph-line transformation of an undirected graph $G$ is another graph $L(G)$ that represents the adjacencies between edges of $G$. $L(G)$ is created in the following manner: for each edge in $G$, create a vertex in $L(G)$; for every two edges in $G$ that have a vertex in common, create an edge between their corresponding vertices in $L(G)$. For directed graphs, which is our case, nodes are considered adjacent in $L(G)$ exactly when the edges they represent form a directed path of length two. A directed path is defined as a path in which the edges are all oriented in the same direction.

\begin{figure}[t!]
    \centerline{\includegraphics[width=10cm]{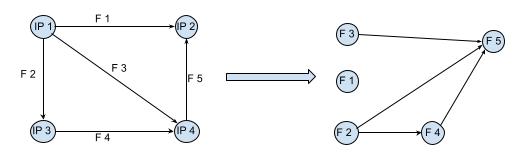}}\vspace{-0.2cm}
    \caption{Illustration of the graph-line transformation}
    \label{fig:graph-line}\vspace{-0.5cm}
\end{figure}

\subsection{Graph Neural Network}
The Graph Neural Network GNN is a specific type of artificial intelligence technique mainly designed to exploit non-structural data, more specifically, graph-based data.
The GNN performs several types of tasks, such as node-level tasks (e.g., node classification), edge-level tasks (e.g., link prediction), and graph-level tasks (e.g., graph classification). Lately, AI researchers have given substantial attention to graph theory in general and graph neural networks in particular, driven by the promising results of employing GNNs in several applications such as molecular analysis, social networks, etc.

The main idea of GNNs is to update each graph node by aggregating the features of the neighbor's features iteratively.
After $K$ iterations, each node is assigned the aggregation of the $K$-hop neighbor's features. So, for example, if $K=2$, each node will have the aggregation of its neighbors and the neighbors' neighbors. 
GNNs are considered outstanding data embedders and feature extractors. They can extract relevant structural information and detect patterns in the graph topology alongside the nodes' features.

\begin{figure*}[t!]
    \centerline{\includegraphics[width=1\textwidth]{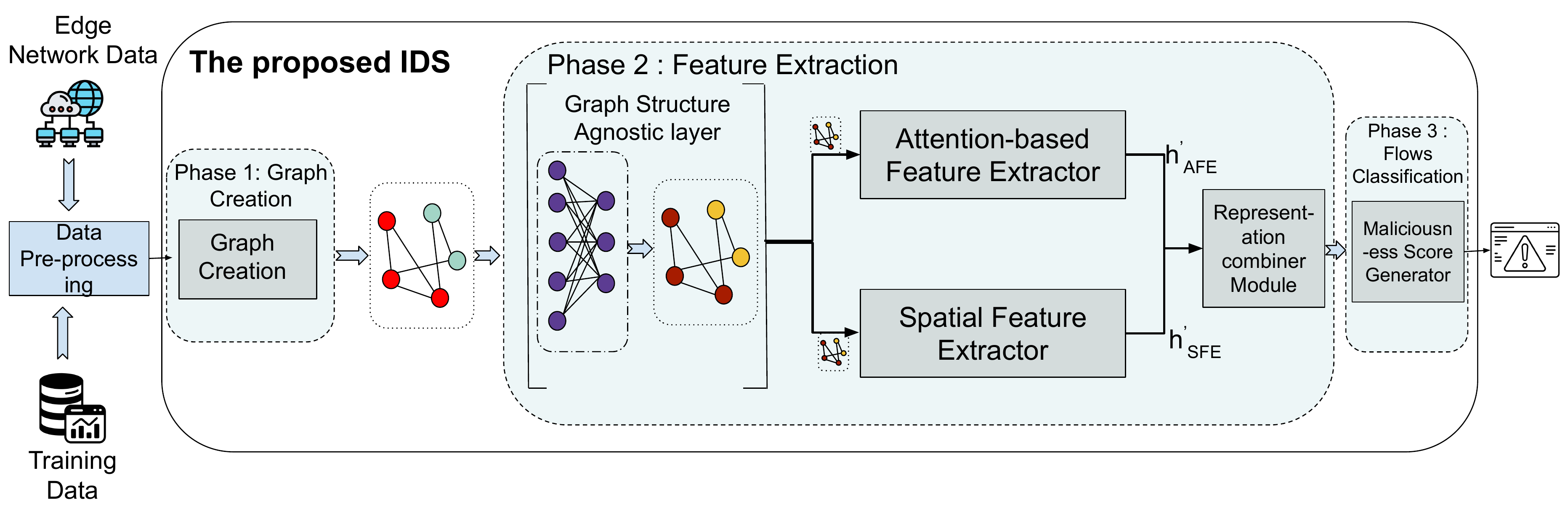}}\vspace{-0.2cm}
    \caption{The proposed GNN-based IDS framework classifying edge communication flows into normal and abnormal behaviors. The model has three main steps. It receives as input network flows to alert the system administrator of any possible attack }
    \label{fig:framework}\vspace{-0.5cm}
\end{figure*}

\section{Proposed Framework}

In Fig. \ref{fig:framework}, we present a flowchart of the proposed IDS framework where we highlight the main steps. The goal is to detect abnormal network behavior and alert the SA to intervene and prevent possible attacks. To this end, data-extraction devices collect information regarding all the communication flows in the network. The collected data contains several types of information that will be used by the model to detect malicious activities. These features provide information about the used protocols, such as the time of the packet connections, the number of original packets, SSL activity, HTTP activity, DNS activity, violation activity, etc.
This information is fed then to the IDS framework, which detects any malicious activity and reports it to the SA administrator.

Prior to the IDS, in the pre-processing phase, the data is processed and transformed into a format consumable by the IDS. The first step of the IDS, namely the graph creation phase, is responsible for processing and transforming the data into a flow-based graph. This graph is structured in a specific manner that will be described later in Section \ref{sec:graph_creation}.
The second step of the IDS exploits the flow-based graph to extract and embed relevant spatial and non-spatial information. In this step, we have three main modules: 
\begin{itemize}
    \item Graph Structure-Agnostic (GSA): this module is responsible for embedding the node attributes (flows-related features) to extract the most relevant information.
    \item Attention-based Feature extractor (AFE): this module exploits the embedded features generated by the GSA layer to aggregate neighbors data while assigning an importance score to each one of them.
    \item Spatial Feature Extractor (SFE): this module extracts the spatial information from the graph using a convolutional graph network GCN.
\end{itemize}
The outputs of AFE and SFE are then combined by the representation combiner and fed to a decision-making module to identify any unusual flows.\\

In the following sections, we explain, in a more detailed manner, each phase of the proposed IDS framework.

\subsection{Data Pre-processing Phase}

Data pre-processing is crucial for all AI algorithms. Indeed, during the training or testing phase, the data must be pre-processed before being forwarded to the model. In our work, the pre-processing module adapts the flows' features to be processable by the GNN-based algorithms. It mainly aims at alleviating the data-related problems that may later cause issues with the model, ensuring more stable learning and obtaining a higher level of accuracy. 

We foremost perform standard normalization of all the features to transform the data to the same scale. This technique prevents the model from prioritizing some features over others.
Afterward, we encode the non-numerical data using one-hot encoding since the majority of AI models do not accept string format categorical features.

In this work, we investigate two datasets: CICIDS 2017 dataset \cite{CICIDS2017} and ToN IoT dataset \cite{9189760}. For instance, both datasets are highly unbalanced, which makes the model biased toward the majority class, the normal flows in our case. The undersampling technique will reduce the number of the normal flows only to have a lesser degree of unbalance for the training and the testing. It is worth mentioning that it is not recommended to perform harsh undersampling and transform the data to be fully balanced. In this latter case, the risk is that the transformed data would no more accurately represent real-world scenarios. Hence, the obtained results would be worthless in practice.

\subsection{Phase 1: Graph Creation}
\label{sec:graph_creation}

The graph creation step consists in transforming a batch of flows into a flow-based graph.
Previous works, such as the ones mentioned in section II \cite{Chang2021GraphbasedSW}, represent the network as a graph where the nodes are the users, each identified with an IP address only or combined with a port number as a second identifier. However, this representation is too straightforward and has several drawbacks. First, using this structure, the problem is transformed into an edge classification task, but network intrusion detection is about capturing the malicious flows and not the attackers. Indeed, detecting users with malicious behavior is more complicated than detecting malicious flows and can be evaded easily. For this, the attackers usually change their IP addresses using virtual private networks.
In addition, the edge classification task is not yet well investigated using GNN theory, and hence, the state-of-the-art results are not encouraging \cite{8954414}. Therefore, recent work in \cite{Chang2021GraphbasedSW} transformed this graph structure into its line-graph representation, where the nodes are transformed into edges and vice-versa, as described in section \ref{sec:graph-line}. However, this line-graph transformation can be computationally costly and does not work well for all graph types (e.g. heterogeneous graphs).\\

In general, the classification between normal and malicious traffic in intrusion detection can be done using the classical representation through two methodologies: 1) apply the graph-line (defined in subsection \ref{sec:graph-line}) transformation to convert the problem to a node-classification task since the flows are classically represented as edges. 2) Classify flows as an edge classification problem. 
Note that the node classification is better developed than the edge classification task in the literature. Hence, it is sometimes preferred to model the problem as node classification to ensure satisfactory performance.\\

In order to improve the performance of graph-based intrusion detection, we seek to find a new graph structure with less complexity than the classical network representation while providing more relevant topological information.
Therefore, we introduce in this paper a novel flow-based graph structure that models the flows as nodes of the graph. The new network representation allows the application of several graph neural networks efficiently. Indeed, having a simple and flow-based structure (nodes represent flows) allows us to avoid graph-line transformation (i.e., graph edge-to-node transformation) or edge classification techniques. \\

The proposed new graph structure aims to model the inter-relation between flows. Thus, we model the network as a weighted graph where each node represents a flow, and the edges are created to link the flows that originated from the same user or are directed to the same user.
Our structure fixes the previously cited issues, enabling the model to attain a high accuracy by providing relevant topological information.

The edges' weights aim to enhance the GNN-based models' pattern recognition by providing information about the two connected nodes (i.e., the two connected flows). For example, our experimental results allowed us to notice that in iterative behavior flows are highly similar, and hence we aimed to assign a similarity score to the edge. Empirically, we chose to assign each edge a weight that consists of cosine similarity score $\zeta_{(u, v)}$ defined as follows:
\begin{equation}
\zeta_{(u, v)} = \frac{u^{T}v}{\|u\|\|v\|} = \frac{\sum_{k=1}^{N_{f}}u_{k}v_{k}}{ \sqrt{\sum_{k=1}^{N_{f}}(u_{k})^{2}}\sqrt{\sum_{k=1}^{N_{f}}(v_{k})^{2}}},
\label{eq:sim_metric}
\end{equation}

where $u$ and $v$ are two vectors representing flows' attributes that contain, respectively, the features $u_k$ and $v_k$ extracted while creating the communication flows. $N_{f}$ is the number of extracted features.
Indeed, attack traffic would likely exhibit a certain similarity among its involved flows, whether these latter are probing traffic, DoS payloads, or bound to a single, propagating attack. \\
The created graph is then forwarded to the feature extraction modules.\\

\begin{figure}[!t]
    \centerline{\includegraphics[width=10cm]{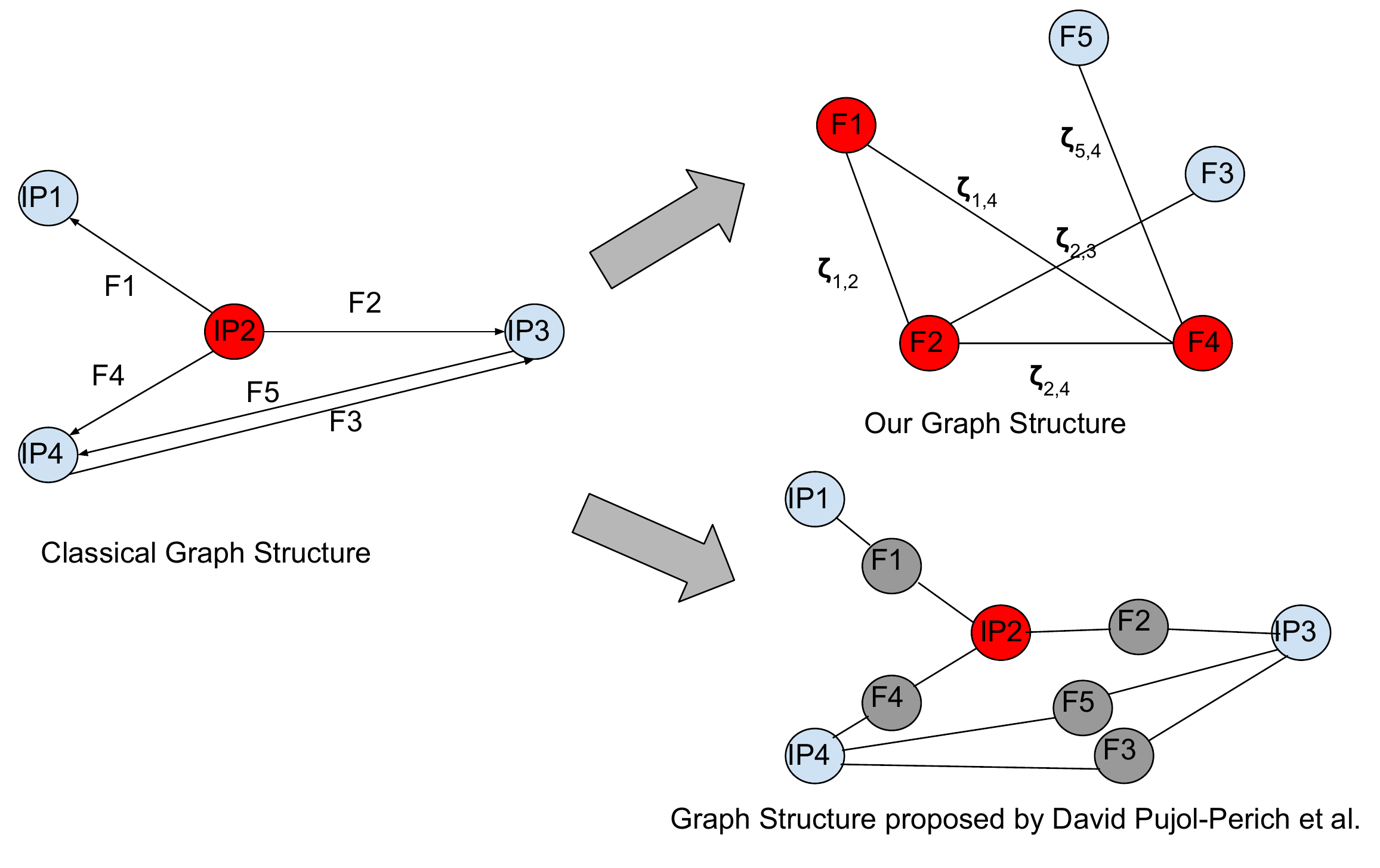}}\vspace{-0.2cm}
    \caption{High level comparison of our flow-based graph structure with classical representation}
    \label{fig:graph_structure}\vspace{-0.5cm}
\end{figure}

\textcolor{black}{
In Fig.~\ref{fig:graph_structure}, we compare our proposed graph representation with the state-of-the-art representations. On the left, we find the classical structure where nodes represent users and edges represent communication flows. On the right, we visualize our structure and the one proposed in \cite{DBLP:journals/corr/abs-2107-14756}.
}

\textcolor{black}{
In the classical representation (the graph on the left), we have one user (IP2 colored in red) who has malicious communication flows with several users (IP1, IP3, and IP4 colored in blue). In our corresponding proposed graph, flows F1, F2, and F4 are interconnected because they originated from the same source user (IP2). Likewise, flows F4 and F5 are connected because they are going to the same user (IP4). The same concept applies to the rest of the flows and weights are assigned to the edges.
The representation proposed in \cite{DBLP:journals/corr/abs-2107-14756}, shows a heterogeneous graph with two types of nodes. The first type characterizes the users (blue nodes) and the second type represents the flows (grey nodes).
}

\begin{figure}[t]
    \centerline{\includegraphics[width=13cm]{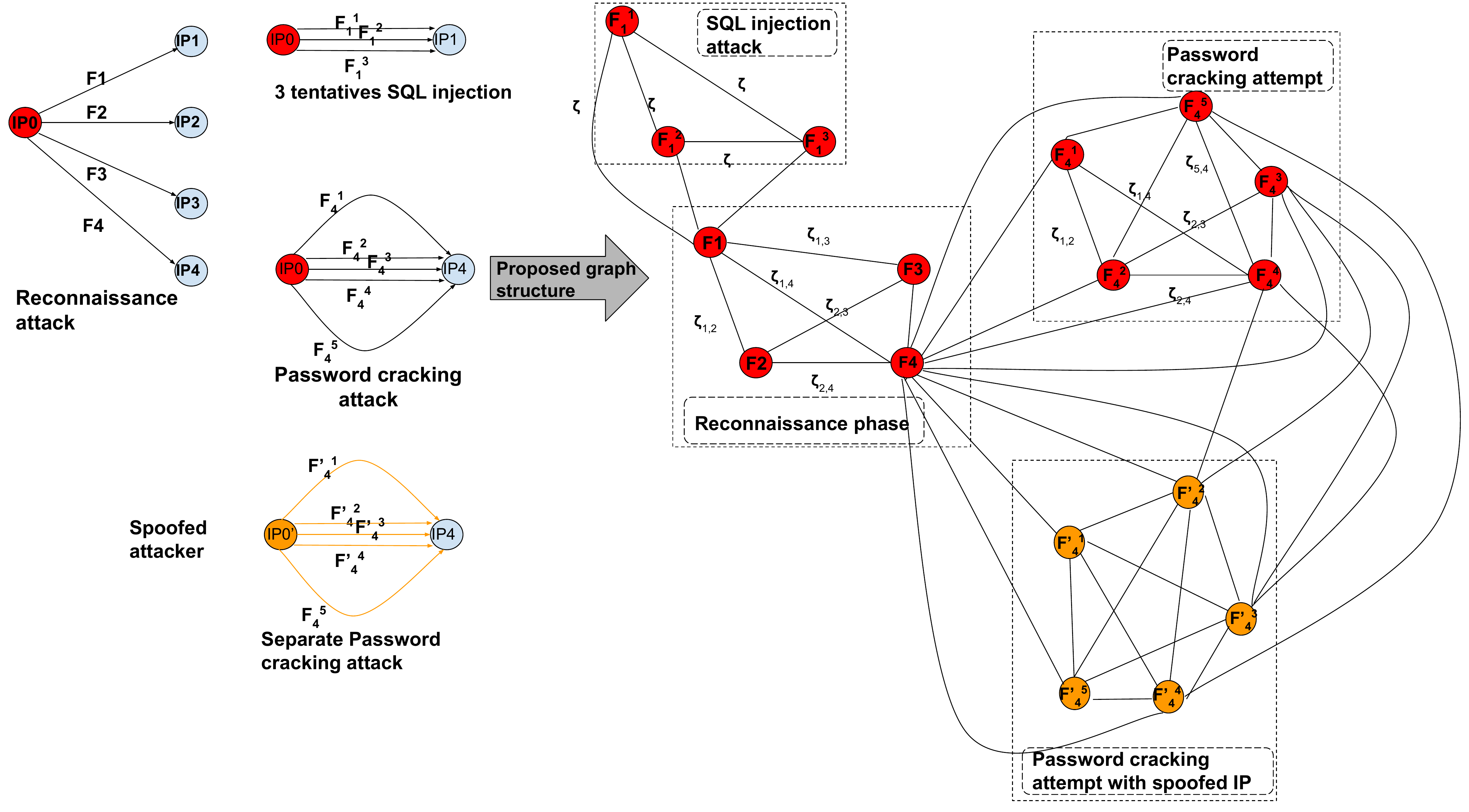}}\vspace{-0.2cm}
    \caption{Illustration of the proposed network representation in case of a multi-step attack. We represent each step of the attack using classical representation (on the left) and the proposed graph representation (on the right). NB: some edges and weights were not presented in the figure for simplicity reasons.}
    \label{fig:complexe_scenario}\vspace{-0.5cm}
\end{figure}

In Figure \ref{fig:complexe_scenario}, we present a multi-step attack scenario, and its equivalent representation using the classical representation (on the left) and our graph structure (on the right).
The attack scenarios depicted in the example of Figure \ref{fig:complexe_scenario} correspond to the following time sequence:\\
$\bullet $ At time $t1$, the attacker IP0 collects information simultaneously on several network devices IP1, IP2, IP3,and IP4. Thus it generates flows $F1, F2, F3,$ and $F4$. This step is represented by the graph entitled "Reconnaissance phase graph". \\
$\bullet $ At time $t2$, the attacker performs an SQL injection attack to exploit a security flaw in one of the IP1 devices. For this, attacker IP0 generates streams $F_{1}^1, F_{1}^2, F_{1}^3$ to target IP1. This step is represented by the graph portion entitled "SQL injection attack".\\
$\bullet $ At time $t3$, the attacker tries to crack a password of user IP4 to gain access to the network. For this, it generates the flows $F_{4}^1, F_{4}^2, F_{4}^3, F_{4}^4, F_{4}^5$ towards the target IP4. This step is represented by the graph entitled "Password cracking attempt".\\
$\bullet $ At time $t4$, the attacker spoofs his own IP address to avoid detection by classical IDS after several unsuccessful connection attempts. Hence, he creates the flows ${F'_{4}}^1, {F'_{4}}^2, {F'_{4}}^3, {F'_{4}}^{4}, {F'_{4}}^5$ towards the target IP4. The portion of the graph that represents the flows after the IP-spoofing is entitled "Password cracking attempt with spoofed IP".

In this scenario, our representation links the spoofed behavior with previous steps of the attack enabling the GNN algorithms' spatial aggregators to link the flows belonging to the same attack or related to the same attack through the clusters created in the graph structure (e.g. SQL injection attack and the password cracking attack). Notice that even after performing IP spoofing, the model can still link the spoofed and pre-spoofed flows.

\begin{itemize}
    \item \textcolor{black}{Relevant topological information about malicious behavior patterns that assists models in distinguishing malicious flows from normal flows accurately.}
    \item \textcolor{black}{Flows-based representation: IP-based solution could be evaded easily using techniques similar to IP spoofing. However, our structure is immune to IP spoofing.}
    \item \textcolor{black}{Relevant information about multi-step attacks which creates a link between each attack step.}
    \item \textcolor{black}{Lower memory consumption in comparison to the solution presented in \cite{DBLP:journals/corr/abs-2107-14756} since it has less nodes and edges. Hence, our solution is more efficient regarding computational power.}
    \item \textcolor{black}{Easier exploitation by the GNN algorithms since the problem is directly formulated as a node classification task. There does not require to pass through graph transformation, such as graph line transformation.}
\end{itemize}

The phases of the proposed framework, described in the subsequent subsections, are created in a way to sufficiently exploit the new graph structure and attain a high precision level.
\subsection{Phase 2: Feature Extraction Phase}
This phase exploits the graph structure and the nodes' features to extract pertinent information to be fed to the decision-making module for the intrusion detection task. The feature extraction is performed using three different modules: the graph structure-agnostic module, the attention-based feature extractor, and the spatial feature extractor. These models are detailed in what follows.

\subsubsection{Graph Structure-agnostic (GSA) Module}:
The nodes' attributes of malicious flows are similar to those of normal flows. Thus, learning a discriminative embedding is required to enable the model to distinguish between normal and abnormal flows.
The graph structure-agnostic module, as the name states, does not rely on the topological information of the graph. Instead, it considers only the flow data to extract a discriminative embedding of ordinary and malicious behaviors.

In this module, the flows' attributes are fed to a Multi-Layer Perceptron (MLP) which embeds the data to a lower space in a data-driven manner and learns a discriminative representation between the extracted data of normal and malicious flows. 
Consequently, the similarity between normal users flows and camouflaged attacker's flows is less significant. Therefore, starting with a discriminative embedding will enhance the model's ability to differentiate between normal and abnormal flows.

\subsubsection{Spatial Feature Extractor (SFE)}:
Graph convolutional network GCN is a GNN that can extract spatial information from graph data. It applies convolution operation on graphs and iteratively updates each node's features by aggregating its neighbors' node and edges features.
This module comprises a two-layer GCN model to convolute on the graph. These layers embed and extract the relevant information in the topology. The convolution calculation process in the GCN is defined as follows:
\begin{equation}
{{\mathbf{H}}^{(l + 1)}} = \sigma \left( {{{{\mathbf{\hat D}}}^{ - \frac{1}{2}}}{\mathbf{\hat A}}{{{\mathbf{\hat D}}}^{\frac{1}{2}}}{{\mathbf{H}}^{(l)}}{{\mathbf{W}}^{(l)}}} \right),
\label{eq:GCN}
\end{equation}
where $\mathbf H^{(l+1)}$ is the $l^{th}$ layer output; $\mathbf{\hat A} \in \mathbb{R}^{n\times n}$ is the adjacency matrix defined as $\mathbf{\hat A} = \mathbf{A} + \mathbf{I}$, with $\mathbf{A}$ the classical adjacency matrix and $\mathbf{I}$ the identity matrix. As a matter of fact, the diagonal elements of $\mathbf{\hat A}$ are equal to 1 to include the investigated node features during the aggregation procedure. 
$\mathbf{\hat D}$ $\in \mathbb{R}^{n\times n}$ is the diagonal node degree matrix of $\mathbf{\hat A}$; $n$ is the size of the nodes set; $\mathbf{W}^{(l)}$ is the trainable weight matrix of the $l^{th}$ layer; $\sigma(\bullet)$ represents the ReLu activation function defined as Relu $= \max(0,\bullet)$. We denote the output of the SFE module  $h'_{SFE}=[h'_1,h'_2,\ldots,h'_N]$ where $h'_i$ indicates the output embedding of the $i^{th}$ entity, and $N$ here is size of the SFE output embedding space.

The two GCN layers learn low-dimensional representations to capture the graph topology, node-to-node relationship, patterns, and other pertinent information about graphs, such as sub-components and vertices.
The classical machine learning models are not capable of extracting such topology-related features, and hence GCN algorithms are more efficient wherever we have graph-structured data.

\subsubsection{Attention-based Feature Extractor (AFE)}

This module exploits the graph topology to extract an embedding of nodes' attributes. This phase is based on the Graph Attention layer (GAT).
The GAT layer takes as input an embedding vector $h= [h_1,h_2,\ldots,h_N]$ and outputs a features embedding vector $h'_{AFE}=[h'_1,h'_2,\ldots,{h'_N}_{emb}]$ where $h_i$ and $h'_i$ indicate the input and output embeddings of the $i^{th} $ entity, respectively, $N$ the size of the input space and $N_{emb}$ the size of the output embedding space. The attention value of an entity can be formalized as follows:
\begin{equation} 
h_{ij} =a(\mathbf{W'}h_{i},\mathbf{W'}h_{j}),
\end{equation}
where $a(\bullet)$ is a mapping function to project the spliced high-dimensional feature to a real value, and $\bold{W'}$ is a linear transformation matrix. The attention value represents the importance of the edge $(h_i,h_j)$, which can be employed to estimate the importance of the head node $h_i$. The attention model learns a weight of attention for each edge and then collects information from neighbors using the calculated priorities as:
\begin{equation} 
h_{i}^{\prime } =\sigma \left({\sum \limits _{j\in \Omega _{i}} {\alpha_{ij} \bold{W'}h_{j}} }\right),
\end{equation}
where $\alpha _{ij}$ represents relative attention weights computed by applying softmax function over all the neighbors' values using the following formula:
\begin{equation} \alpha _{ij} =\textrm{softmax}_{j} (h{_{ij}})=\frac {\exp(h_{ij})}{\sum \limits _{n\in \Omega _{i}} {\sum \limits _{r\in \Re _{in}} {\exp(h_{in})}}}\end{equation}
where $\Omega _{i}$ denotes the neighbors set of nodes $h_i$, $\Re _{in}$ denotes the relations set which connects between $h_i$ and $h_n$.

The GAT's main goal is to assign importance to different neighboring nodes rather than giving an analogous weight to all of them. This concept is essential for the detection of intrusions. In fact, it allows the models to detect unusual flows where the user conceals its malicious activity by building several normal communications with normal users. At this level, the similarity score is used to provide the model with more information about the similarity of the flows.

\subsection{Phase 3: Flows Classification}
The decision-making and alerting functionalities of the proposed IDS are performed in this module. The combined embedding of the spatial and non-spatial information is used to calculate a maliciousness score $P_m$. The higher $P_m$ is, the likelier the investigated flow is a potential attack.
The $P_m$ is calculated using a multi-layer perceptron MLP network, precisely a 3-layer MLP, trained to distinguish between the normal and malicious flows from the previously combined data.
Each layer of the MLP is defined as follows:

\begin{equation}Z_{l}=f(\bold{W_{m}}^{l}Z_{l-1}+b_{l}+\varpi_{e}^{l}e),\end{equation}
where $\bold{W_{m}}^{l}$ and $b_l$ are respectively the weight and bias in the $l^{th}$ layer, $e$ is the re-construction error, $\varpi_{e}^{l}$ its corresponding weight, and $f$ is the activation function which typically a non-linear function such as the sigmoid, ReLU, or tanh. After the computation of $P_m$, any flow with a score $P_m > S$ is considered an attack, with $S$ a given threshold score.
$S$ is a crucial parameter for controlling the sensitivity of the proposed IDS. Indeed, the higher $S$ is, the more sensitive the IDS is to any abnormal behavior, and hence the higher the false-positive ratio as well.

\section{Evaluation Procedure}

In this section, we present and analyze two of the most used datasets in previous works to assess the quality of their proposed IDS solutions. We also discuss the existing evaluation approaches and highlight their drawbacks.
Two major datasets are exploited in the literature:
\begin{itemize}
    \item CICIDS 2017 dataset~\cite{CICIDS2017}: CICIDS2017 is a labeled network flows dataset alongside a full packet version in PCAP format. It covers the most common network attacks such as DoS, HULK, DDoS, FTP-Patator, DoS Slowloris, and SSH-Patator attacks. CICIDS 2017 is created by capturing five days network stream. CICIDS 2017 flow-based version is created using the CIC-FlowMeter \cite{CICIDS2017}.
    \item ToN IoT dataset~\cite{9189760}: ToN IoT is a heterogeneous dataset released in 2020 by the Intelligent security group UNSW Canberra, Australia. The dataset contains a huge number of realistic attack scenarios such as scanning, DoS, Ransomware, and Backdoor attacks. Ton IoT network flow-based version is created using the NetFlow.
\end{itemize}

By reviewing the literature on these datasets, we find some flaws with the creation procedure of the CICIDS 2017 dataset. For example, the authors in \cite{EngelenGints2021TaID} stated that the creation procedure has some issues in the CICFlowMeter tool that violate the correct implementation of network traffic. Moreover, they mentioned that CICFlowMeter suffers from an inaccurate labeling strategy that caused several mislabeled flows, which could mislead the learning of ML models. The ToN IoT dataset is more recent, and no previous work highlights any problem that could affect the performance of machine learning models.

We have conducted a thorough analysis of both datasets. For the CICIDS 2017, we notice that elementary models, like the decision tree-based model, can perfectly distinguish malicious flows. Moreover, these models have relatively good validation metrics since their first iterations, which is unusual in the learning process of ML models. Furthermore, we notice that during training, the validation metrics are higher than the training metrics. This behavior of ML models alerts on significant issues with the dataset or the validation methodology.

In addition, for the CICIDS 2017, we observe a high similarity between flows of the same user. We explain this by the improper termination of TCP sessions as explained in \cite{EngelenGints2021TaID}. Indeed, due to these issues, a single attack could be segmented into several flows, and evidently, these segments would be highly correlated. The ToN IoT is also slightly affected by this issue. It is tough to avoid such a problem while transforming the packet-based dataset into the flow-based one.
Nevertheless, the problem is boosted in the case of CICIDS 2017 since the distribution of malicious flows per attacker is highly unbalanced. This means that some attackers are responsible for most of the attacks. 

Table \ref{tab:stats_dataset} presents the number of flows labeled as malicious per attacker for both datasets. We notice that the distribution of attackers is not balanced for the case of CICIDS 2017, and we have one single IP address (the firewall IP address) assigned to the four exterior IP addresses. Hence, this is a massive issue in the graph structure because one node represents all the external users. More specifically, in the CICIDS 2017, all the flows come from the external network (.i.e, users with IP addresses: 205.174.165.73, 205.174.165.69, 205.174.165.70, 205.174.165.71) are assigned the 127.16.0.1 IP address.
The ToN IoT dataset has better distributed malicious flows per attacker, which is clearly observed in Table \ref{tab:stats_dataset}.

\begin{table}[!tb]
    \caption{Distribution of flows labeled as attacks per each attacker IP in the TON IoT and the CICIDS 2017. NB: The ToN IoT contains 19 attackers, but we presented the main 10 attackers. The 127.16.0.1 IP address in the CICIDS 2017 belongs to the firewall. It was assigned to all packets coming from the external network, specifically the 4 users with the following IP addresses: 205.174.165.73, 205.174.165.69, 205.174.165.70, 205.174.165.71}\vspace{0.5cm}
    
    \begin{minipage}{.5\linewidth}
      \centering
\begin{tabular}{cc}
\toprule
      & \multicolumn{1}{c}{\hspace{-2cm} ToN IoT}\\ \hline
  \multirow{2}{*}{}     IP Address & \textnumero of flows         \\
  \midrule

    192.168.1.192   & 23047      \\     \midrule

    192.168.1.30    & 61585   \\   \midrule

  192.168.1.31 & 30355    \\   \midrule
  192.168.1.32  & 27227     \\   \midrule
    192.168.1.33  & 9439    \\   \midrule
  192.168.1.34   & 528    \\   \midrule
    192.168.1.36   & 631      \\     \midrule
    192.168.1.37  & 7500    \\   \midrule

    192.168.1.38   & 10    \\   \midrule

    192.168.1.39    & 692   \\  

 \bottomrule
\end{tabular}
\label{tab:Multi_nav}
    \end{minipage}%
    \vline
    \begin{minipage}{.5\linewidth}
      \centering
\begin{tabular}{cc}
\toprule
      & \multicolumn{1}{c}{\hspace{-2cm} CICIDS 2017}\\ \hline
  \multirow{2}{*}{}     IP Address & \textnumero of flows         \\
  \midrule
172.16.0.1  &  554561       \\   \midrule
  192.168.10.12 & 2    \\   \midrule
  192.168.10.14  & 209     \\   \midrule
    192.168.10.15  & 366    \\   \midrule
  192.168.10.17   & 2    \\   \midrule
    192.168.10.5   & 179      \\     \midrule
    192.168.10.50  & 3    \\ 
  \midrule
    192.168.10.8   & 307    \\   \midrule

    192.168.10.9    & 226   \\   \midrule
 205.174.165.73     & 701    \\ \bottomrule
\end{tabular}

    \end{minipage} 
    
\label{tab:stats_dataset}
\end{table}

The classic evaluation procedure that randomly splits users' flows between training and testing sets is not recommended in this situation. In fact, due to the similarity between the flows of the same user and the segmentation of a single communication into multiple flows, we can conclude that this procedure is biased by seeing correlated data in the training and testing. To ensure reliable validation, we evaluate our model using an IP-based evaluation procedure. The idea consists in splitting the datasets into train and test sub-data using the distribution of malicious flows per attacker and the type of attacks. Consequently, we obtain separate train and test data regarding the investigated users and the type of attacks. More precisely, we have a list of training IP addresses and another different one for the testing, where the attack types in the training set differ from those in the testing set. This methodology fixes the issue of flows' similarity and validates the capability of the model to detect new types of attacks. Indeed, the model will be evaluated on new IP addresses with unseen types of attacks.

\section{Results \& Discussion}
\subsection{Experimental Settings}
We have implemented the proposed framework in a python environment. The Pytorch Geometric and Netwrokx libraries are used to develop the proposed GNN-based algorithms. The training and testing are performed using an Nvidia GTX 1080 Ti graphic card. Table \ref{tab:settings} presents the parameters required to rebuild our proposed work. The model comprises 2 MLP layers for the discriminative layers, 2 GCN layers, and one GATv2 layer. The score generation is performed using a 3-layer MLP.
For the training, we employ the Cross-Entropy Loss as the performance measure of the classification problem 
\cite{DBLP:journals/corr/abs-2107-14756} and Adam as the parameters' optimizer.

\vspace{-0.5cm}
\begin{table}[h!]
\centering
\caption{Development Environment Settings}
\addtolength{\tabcolsep}{-2pt}\begin{tabular}{>{\centering\arraybackslash}p{1.9cm}
>{\centering\arraybackslash}p{1.5cm}
>{\centering\arraybackslash}p{1.5cm}
>{\centering\arraybackslash}p{1.5cm}
>{\centering\arraybackslash}p{1.5cm}
>{\centering\arraybackslash}p{1.5cm}
>{\centering\arraybackslash}p{1.5cm}
>{\centering\arraybackslash}p{1.5cm}
}

\toprule
\multicolumn{1}{c}{\textbf{Learning rate}}   & \textbf{Adam's betas} & \textbf{Adam's eps} & \textbf{Adam's weight decay} & \textbf{Pytorch version} & \textbf{CUDA version} & \textbf{PYG version} \\ \midrule
\multirow{1}{*}  0.0001        & 0.98    & 1e-8 & 1e-6  & 1.11.0 & 11.1& 1.9    \\

\bottomrule
\end{tabular}
\label{tab:settings}
\end{table}

\begin{figure}[t!]
\centering
    \fbox {\includegraphics[width=10.5cm, height= 8cm]{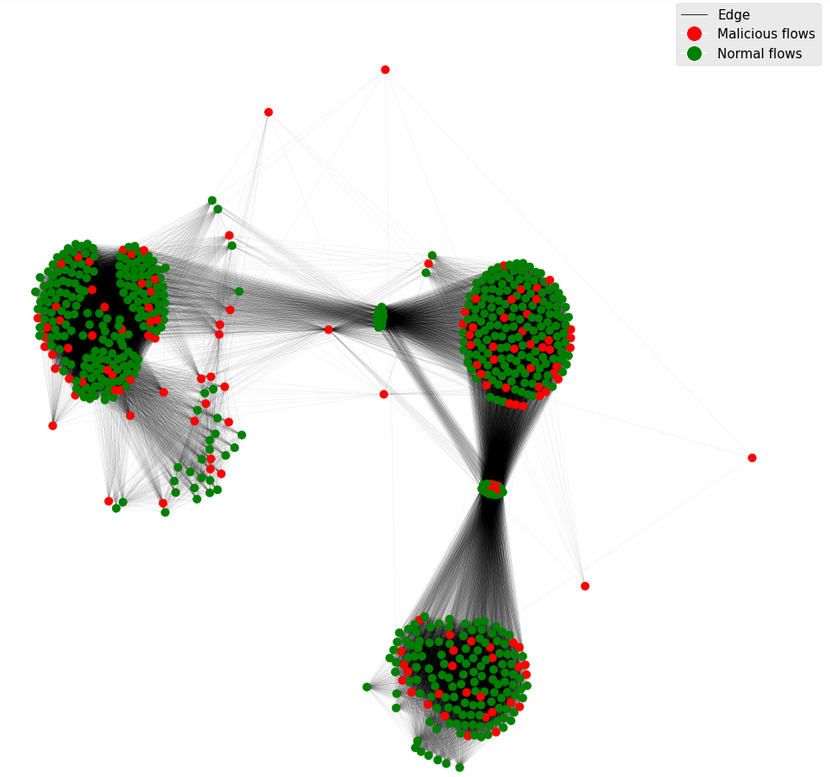}}
    \caption{Illustration of the proposed graph structure created from a sample of ToN IoT dataset's flows}
    \label{fig:zoom_network}
\end{figure}
\subsection{Evaluation Metrics}
In order to show the efficiency of the proposed models, we use several metrics that provide several evaluation aspects. To calculate these metrics, we define the true positive value as the model's output, where it correctly predicts the malicious class. Similarly, a true negative is where the model correctly predicts the benevolent class. On the other side, the false positive is a model's outcome where the model incorrectly predicts the malicious class, and a false negative is where the model incorrectly predicts the benevolent class.
The metrics used during the evaluation are defined as follows:

- Precision: The ratio of correctly predicted positive observations to the total predicted positive observations.
\begin{equation*} Precision=\frac{True\ Positive}{True\ Positive+False\ Positive}.\end{equation*}

- Recall (Sensitivity): The ratio of correctly predicted positive observations to all observations in the actual class.
\begin{equation*} Recall=\frac{True\ Positive}{True\ Positive+False\ Negative}.\end{equation*}

- F1-score: The harmonic mean of Precision and Recall. In other words, it considers both false positives and false negatives into account.
\begin{equation*} F1\ Score=2 \times \frac{(Recall\times Precision)}{(Recall+Precision)}.\end{equation*}

- Area-Under-Curve (AUC): The area between the ROC curve (.i.e the curve plotting the true positive rate vs. the False Positive Rate (FPR)) and the x-axis. Indeed, it is a measure of the ability of a classifier to distinguish between classes. The higher the AUC, the better the model distinguishes between the classes.

\subsection{Results}

This section presents the numerical and graphical results to illustrate the performance of our intrusion detection framework and highlight its efficiency.
\textcolor{black}{First, we illustrate, in Fig. \ref{fig:zoom_network}, a visualization of a flows subset taken from the Ton IoT dataset using our proposed representation described in subsection \ref{sec:graph_creation} (i.e. weights are not presented in the Figure).
In this Figure, we have nodes colored in red that represent the malicious flows in the network and nodes colored in green that represent normal flows.
The graph of flows provides several spatial information, including clusters, sparse nodes, inter-cluster relations through nodes, etc. The spatial information does not have a unique explanation. In other words, the cluster, for example, could be created if a group of flows are included in the same communication, in different communications but related between each other (e.g. distributed attacks), or in successive related communication (e.g. multi-step attacks), etc.
Moreover, we can notice the presence of nodes that link two clusters, and their flows are related to several flows from different clusters. The creation of these spatial components can be understood from the discussion of Fig. \ref{fig:complexe_scenario}. \\
The GNN algorithms exploit the statistical attributes attached to neighboring flows alongside the previously mentioned spatial information for learning attack patterns. Indeed, there are several patterns in the graph, and the model is responsible for learning and detecting these patterns using the topological information and flows attributes.}

\begin{table}[t!]
\centering
\caption{High-level Runtime Performance Comparison }
\addtolength{\tabcolsep}{2pt}\begin{tabular}{>{\centering\arraybackslash}p{5.2cm}
>{\centering\arraybackslash}p{2.2cm} 
>{\centering\arraybackslash}p{2.2cm}
>{\centering\arraybackslash}p{1.8cm}
>{\centering\arraybackslash}p{1.8cm}
>{\centering\arraybackslash}p{1.8cm}}

\toprule
\multicolumn{1}{c}{\textbf{Metrics}} & \textbf{Memory Consumption (MB)}  & \textbf{Energy Consumption (Kw/h)} & \textbf{Processing time (s)}  \\ \midrule
\multirow{1}{*}{Our network representation} &  11.3      &  \bf 0.02  &  \bf 3.14   \\\midrule
\multirow{1}{*}{Classical network representation}  &  \bf 9.17        & 0.04 &  3.7    \\
\midrule
  \multirow{1}{*}{ Network representation proposed by \cite{DBLP:journals/corr/abs-2107-14756} }   &  17.6  &  0.05 &    4.2  \\

\bottomrule
\end{tabular}
\label{tab:runtime_perf}
\end{table}

In Table \ref{tab:runtime_perf}, we conducted a performance analysis from memory and energy consumption point of view. The results confirmed the aforementioned characteristics of our network structure. The Table \ref{tab:runtime_perf} proves that our structure outperforms the classical graph structure in terms of energy consumption and the required time to process the structure in order to make it consumable by the GNN algorithms.
On the other hand, the classical network representation is the one that ensures the lowest level of memory consumption, but it does not provide relevant features in comparison to our graph representation or the one proposed by \cite{DBLP:journals/corr/abs-2107-14756}.

\begin{figure}[t!]
\centering
\begin{subfigure}{.5\textwidth}
  \centering
  \includegraphics[width=1.1\linewidth]{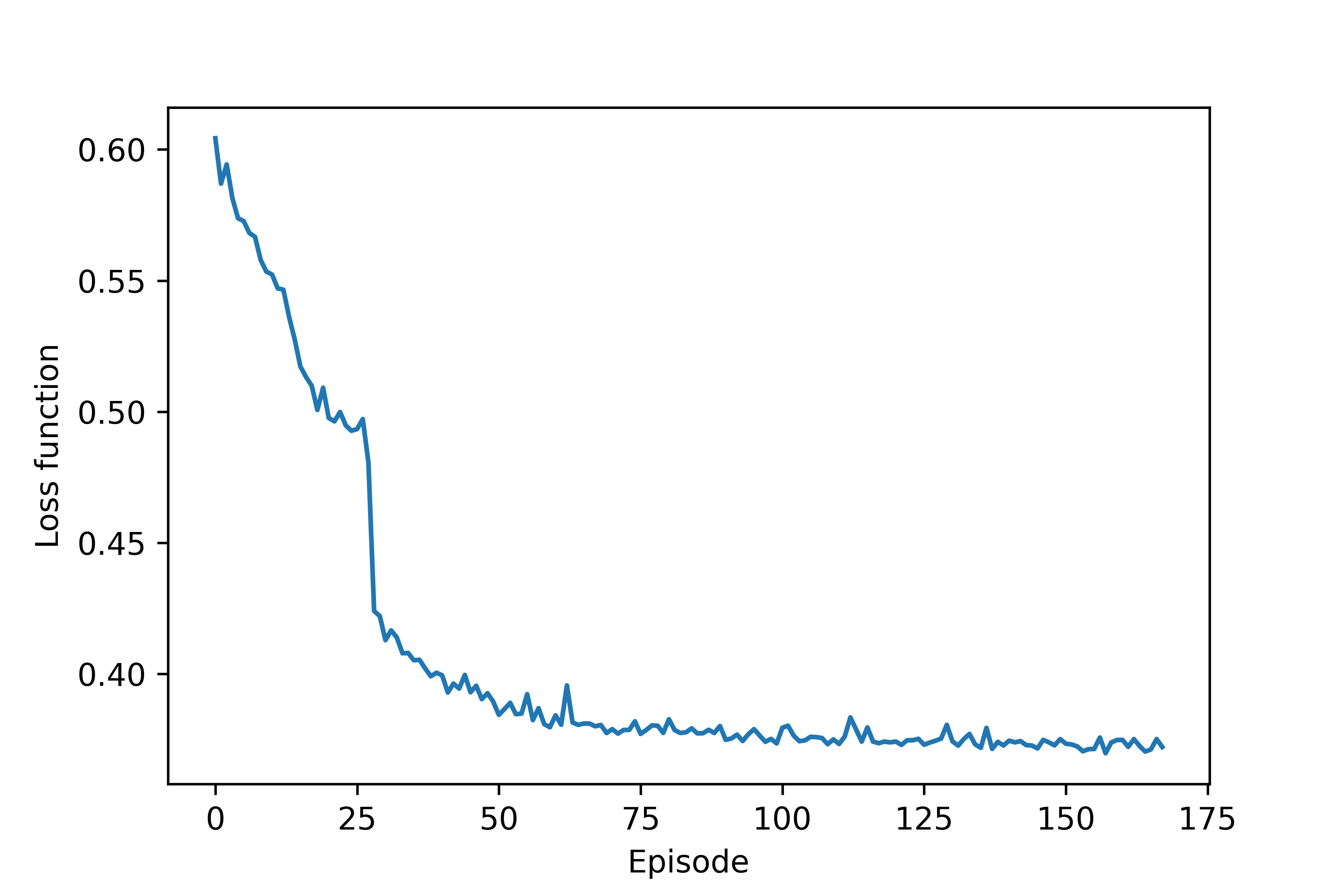}
  \caption{Training loss evolution}
  \label{fig:training_loss}
\end{subfigure}%
\begin{subfigure}{.5\textwidth}
  \centering
  \includegraphics[width=1.1\linewidth]{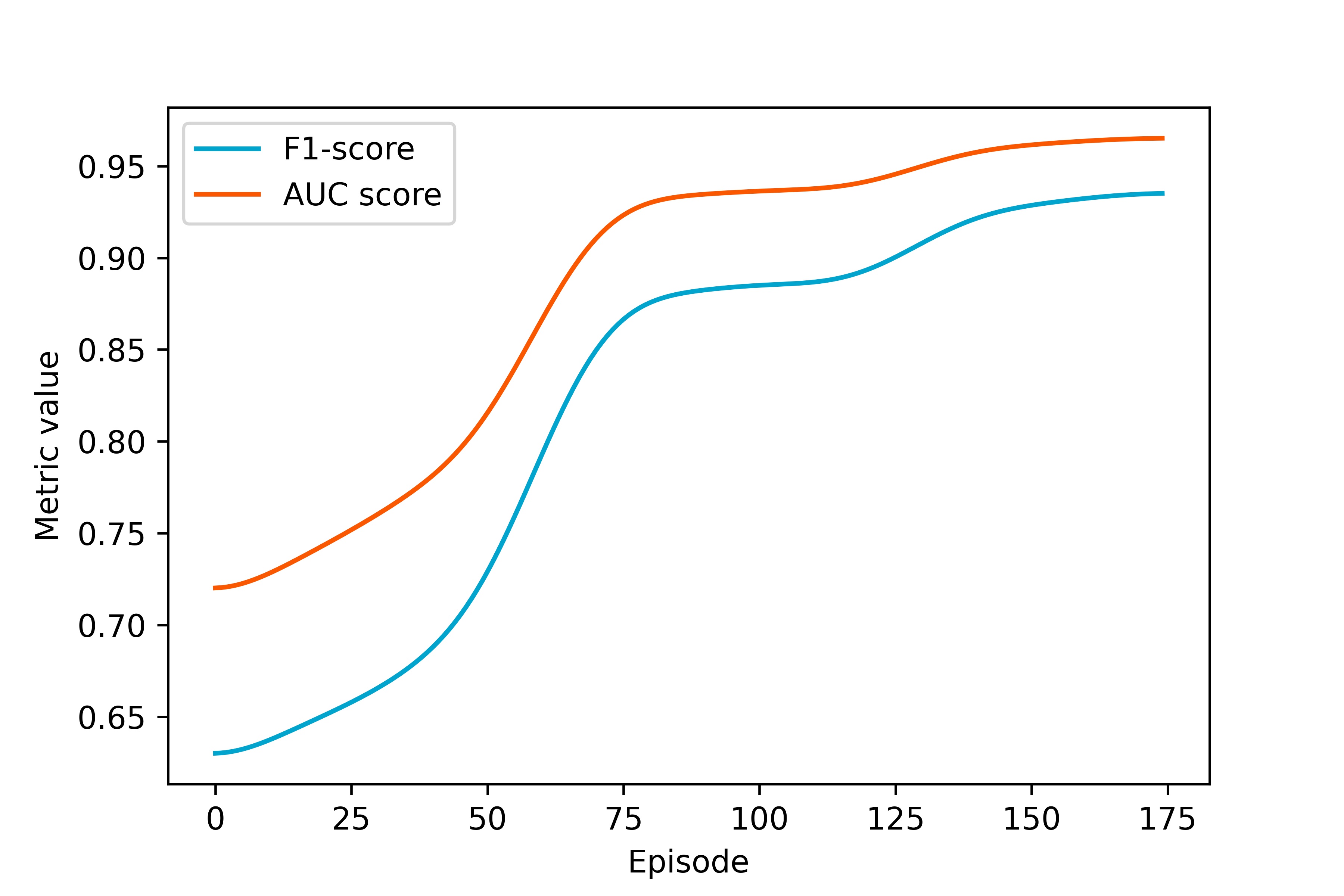}
  \caption{Evaluation metrics evolution}
  \label{fig:metrics_evaluation}\label{fig:sub2}
\end{subfigure}
\caption{The evolution of training loss, and two validation metrics (F1-score and Area-Under-Curve}
\label{fig:tranining_evolution}
\end{figure}

Fig. \ref{fig:tranining_evolution} shows the evolution of the loss function, the F1-score, and the AUC score during the framework training. Fig. \ref{fig:training_loss} presents the cross-entropy loss function. The loss decreases until it converges after almost 100 iterations. This smooth decrease in the curve demonstrates the stability of the model's training. Meanwhile, in Fig.~\ref{fig:metrics_evaluation}, the validation metrics F1-score and AUC increase to reach 93.7 and 96.6, respectively.\\

\textcolor{black}{To endorse the graphical results and compare our model performance with other benchmark models, we have trained and tested several state-of-the-art models using the described IP-based evaluation. Specifically, an MLP-based model \cite{9001867}, an ML-based model \cite{9457024} (.i.e., XGboost-based model), and two GNN-based models \cite{DBLP:journals/corr/abs-2103-16329,DBLP:journals/corr/abs-2107-14756}. The results of this comparison are presented in Table \ref{tab:evaluatio_metric}. The table exhibits several metrics \cite{9937801} used to compare the investigated models' performance to our proposed framework applied to the graph representation with/without the similarity weights $\zeta$.  Our model outperforms the previously stated models in terms of F1-score and false positive rate. The high F1-score of our model proves its capability to accurately classify the flows, and the FPR rate demonstrates its ability to distinguish between normal and malicious flows. The model performs better when applied on a weighted graph than without the weights, highlighting that $\zeta$ provides relevant information about attacks, specifically, iterative malicious behavior.}

 In addition, in Table \ref{tab:evaluatio_metric}, the ML-based model \cite{9001867} reaches a higher recall metric compared to our framework. Indeed, the recall metric is defined as the ratio of correctly predicted positive observations to all observations in the actual class. The precision metric is the ratio of correctly predicted positive observations to the total predicted positive observations. In other words, the recall metric is penalized whenever a false negative is predicted, and precision is penalized whenever a false positive is predicted. Thus, the more we increase the precision, the more we reduce the false positive rate. Conversely, a higher recall means that our machine learning model will have a lower false negative rate.
 The model proposed in \cite{9001867} focused on reducing the false negative rate only during the learning phase, and hence it has a high false positive rate. This model should be avoided in practice since having a high FPR will cause tremendous computational power and human resources (.i.e cybersecurity engineers' efforts), which can be wasted on irrelevant alerts.

\begin{table}[t!]
\centering
\caption{IP-based numerical validation}
\addtolength{\tabcolsep}{2pt}\begin{tabular}{>{\centering\arraybackslash}p{3.5cm}
>{\centering\arraybackslash}p{1.5cm}
>{\centering\arraybackslash}p{1.5cm}
>{\centering\arraybackslash}p{1.5cm}
>{\centering\arraybackslash}p{1.5cm}
>{\centering\arraybackslash}p{1.5cm}
>{\centering\arraybackslash}p{1.5cm}
>{\centering\arraybackslash}p{1.5cm}}

\toprule
\multicolumn{1}{c}{\textbf{Metrics}} & \textbf{F1-score}  & \textbf{AUC} & \textbf{Recall} & \textbf{Precision}& \textbf{FPR}  \\ \midrule
\multirow{1}{*}{Our work without $\zeta$} &  0.915        & 0.954   & 0.992 & 0.848 & 0.083   \\\midrule
\multirow{1}{*}{Our work with $\zeta$}  &  \bf 0.937        & \bf 0.965 & 0.991 & \bf 0.886 & \bf 0.057    \\
\midrule
   \multirow{1}{*}{E-graphsage model \cite{DBLP:journals/corr/abs-2103-16329}}   &  0.88  &  0.92 &   0.962 & 0.82 & 0.087  \\  \midrule
\multirow{1}{*}{GNN-based model \cite{DBLP:journals/corr/abs-2107-14756}}   &  0.902   &  0.932 &  0.981 & 0.835 & 0.093    \\ \midrule
 \multirow{1}{*}{ML-based model \cite{9001867}}   &  0.4807       &  0.7693 &  \bf 0.9957 & 0.3168  & 0.14   \\ \midrule

 \multirow{1}{*}{MLP-based Model \cite{9457024}}   &   0.3438       &  0.8225& 0.9483 & 0.3622 & 0.12  \\

\bottomrule
\end{tabular}
\label{tab:evaluatio_metric}
\end{table}


 


\textcolor{black}{The presence of data leakage issue is clearly seen when performing an evaluation of several models using the classical validation procedure. To lucidly show the presence of the issue, we select several models in a manner to have a basic model (i.e., XGBoost model), a deep learning model (i.e., MLP-based model), and two advanced models (i.e., GNN-based models). For each model, we calculate the F1-score on the testing subset that contains flows chosen randomly from the ToN IoT dataset. 
The results showed that all the evaluated models reached high F1-scores. For example the model proposed in \cite{Chang2021GraphbasedSW,DBLP:journals/corr/abs-2107-14756} reached respectively $0.998$ and $0.99$ , the two benchmark models described in~\cite{9001867,9457024} achieved respectively, $0.996$ and $0.989$. }
\textcolor{black}{As a matter of fact, the high performance of all these models indicates either the triviality of the intrusion detection task, which is clearly not true, or an inappropriate evaluation procedure. The latter case seems more likely as the task is considered one of the most burdensome tasks in the cybersecurity field, and a high level of accuracy is hard to reach.}

\begin{figure}[!t]
\centering
\begin{minipage}{.5\textwidth}
  \centering
  \includegraphics[width=\linewidth]{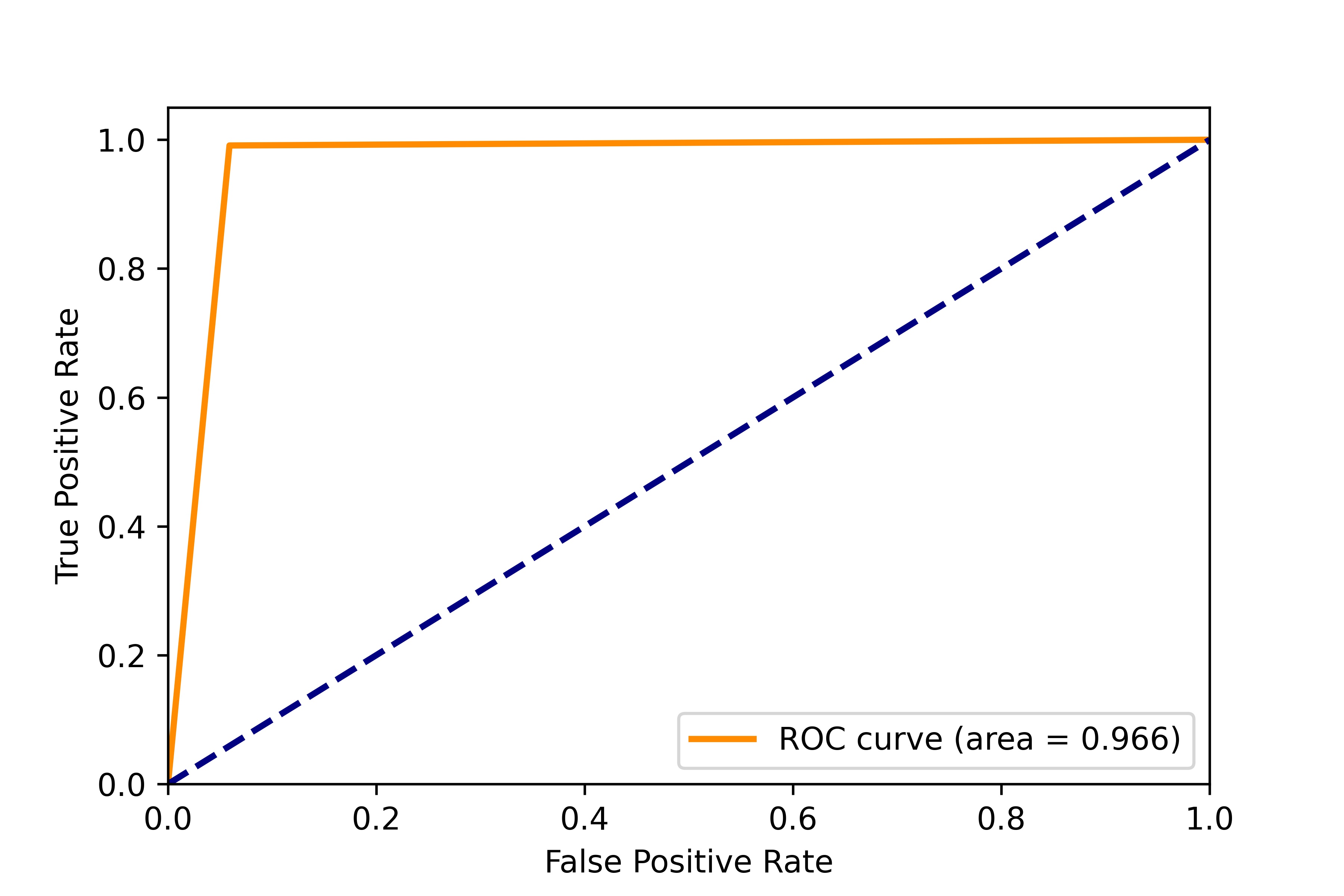}
  \captionof{figure}{Receiver Operating Characteristic (ROC) curve}
  \label{fig:roc}
\end{minipage}%
\begin{minipage}{.5\textwidth}
  \centering
  \includegraphics[width=\linewidth]{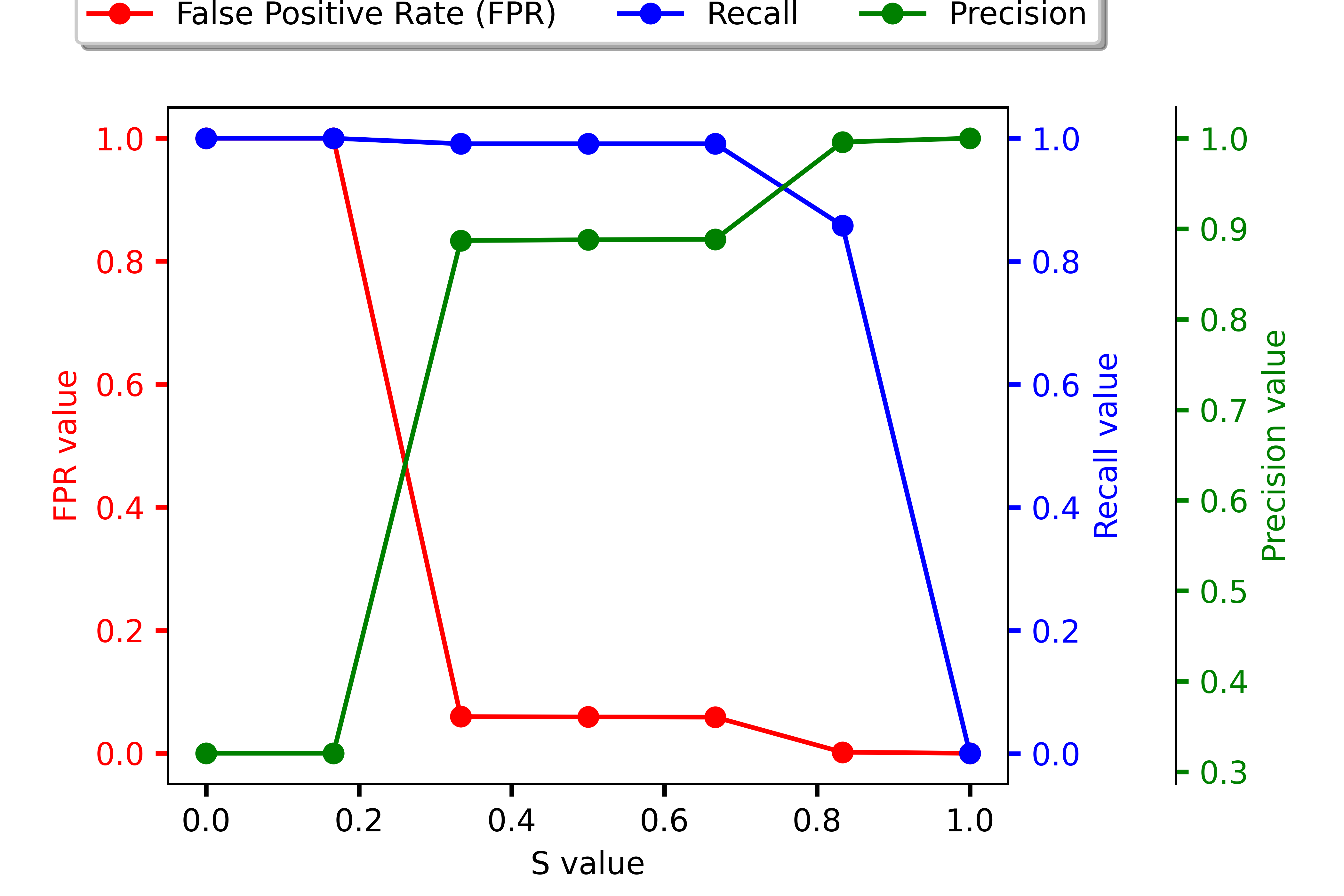}
  \captionof{figure}{Illustration of  $S$ value effect}
  \label{fig:S_value_effect}
\end{minipage}
\end{figure}

On the other hand, the results of the IP-based splitting methodology presented in Table \ref{tab:evaluatio_metric} show that our model achieves a precision of 88.6\%, meaning that 88.6\% of the framework's alerts raised to the SA is correct.
In addition, the recall value of 99.1\% shows the capability of our framework to detect 99.1\% of the attacks correctly.
The AUC metric of 96.6\%, calculated from the ROC curve presented in Fig. \ref{fig:roc}, confirms that our proposed architecture (i.e., GSA module, AFE, and similarity metric) enables the model to differentiate between normal and malicious flows, even for those that belong to camouflaged users.

We recommend using the IP-based splitting methodology over the random splitting for the intrusion detection task. The previous results show that the IP-based validation technique is more reliable than the classical one. Moreover, the IP-based evaluation procedure consists of choosing training and testing IP addresses to have different attacks in both training and testing sets. Consequently, it evaluates the model's ability to recognize new attack patterns.
Accordingly, the achieved results of our framework, exploiting the proposed graph structure, using the IP splitting evaluation prove that our model is capable of detecting new attack patterns that were not included in the training phase.\\

Finally, Fig.~\ref{fig:S_value_effect} illustrates three evaluation metrics evolution as a function of the sensitivity parameter $S$. We notice that for a small value of $S$ we have a high recall, a high false-positive rate, and a low precision. If $S$ is high, we obtain high precision, a low false-positive rate, and a low recall.
To ensure a high precision, a low-false positive rate, and a high recall, $S$ should be in the range $[3.5, \ldots, 6.5]$. At this range, we obtain a 0.057 as a false positive rate FPR.
Note that the SA changes the parameter $S$ to control the sensitivity of the IDS. For example, the SA can choose to increase $S$ if they think an attack is happening and want to detect any miniature anomaly in the network. Thus, the FPR will increase, but the SA will be sure to detect any abnormal activity.


\section{Conclusion}

\textcolor{black}{In this paper, we have proposed a novel flow-based graph structure to represent network communications. This graph structure provides relevant topological information regarding attackers' behavioral patterns, iterative malicious behavior, the link between successive and parallel attacks, etc. Our proposed GNN-based intrusion detection framework manipulates this structure to learn a generalization of malicious behavior patterns. The created graph is exploited in the first step by a graph structure agnostic module responsible for embedding the nodes' features. This embedding is performed by learning a discriminative representation of the flows' features in order to maximize the dissimilarity between the normal and malicious flows. Then, two GNN-based feature extractors extract relevant topology-related information by exploiting the graph attention mechanism and graph convolutional network. The extracted features are combined and forwarded to a decision-making module responsible for detecting potential attacks.
To ensure an efficient evaluation, we have analyzed the existing validation methodologies and operated an IP-based evaluation procedure that is more representative of practical IDS scenarios.
We have conducted experiments that quantify the framework's accuracy in distinguishing between normal and anomalous flows. Finally, we have compared our model to other ML-based and Graph-based solutions, showing that it outperforms these existing solutions. 
In future work, we plan to enhance the graph structure to cover any potential information loss and to provide more information about specific attacks. Furthermore, we will extend this IDS to its distributed version in order to improve energy efficiency, memory consumption and reduce ML footprints.}

\section{Acknowledgement}
The present work has received funding from the European Union’s Horizon 2020 Marie Skłodowska Curie Innovative Training Network Greenedge (GA.No.953775).

\bibliographystyle{splncs04}
\bibliography{references}

\begin{thebibliography}{10}
\providecommand{\url}[1]{\texttt{#1}}
\providecommand{\urlprefix}{URL }
\providecommand{\doi}[1]{https://doi.org/#1}

\bibitem{9189760}
Alsaedi, A., Moustafa, N., Tari, Z., Mahmood, A., Anwar, A.: {TON IoT Telemetry Dataset: {A} New Generation Dataset of IoT and IIoT for Data-Driven Intrusion Detection Systems}. IEEE Access  \textbf{8} (Sept 2020). \doi{10.1109/ACCESS.2020.3022862}

\bibitem{9937801}
Brahim, S.B., Ghazzai, H., Besbes, H., Massoud, Y.: A machine learning smartphone-based sensing for driver behavior classification. In: 2022 IEEE International Symposium on Circuits and Systems (ISCAS). pp. 610--614 (May 2022). \doi{10.1109/ISCAS48785.2022.9937801}

\bibitem{Chang2021GraphbasedSW}
Chang, L., Branco, P.: {Graph-based Solutions with Residuals for Intrusion Detection: the Modified E-GraphSAGE and E-ResGAT Algorithms}. ArXiv  \textbf{abs/2111.13597} (2021)

\bibitem{Dubey2013ASO}
Dubey, R., Pathak, P.N.: A survey on anomaly and signature based intrusion detection system({IDS}). International Journal of Managment, IT and Engineering  \textbf{3},  334--354 (2014)

\bibitem{EngelenGints2021TaID}
Engelen, G., Rimmer, V., Joosen, W.: {Troubleshooting an Intrusion Detection Dataset: the CICIDS2017 Case Study}. pp. 7--12. IEEE (May 2021)

\bibitem{9527927}
Garrido, J.S., Dold, D., Frank, J.: {Machine learning on knowledge graphs for context-aware security monitoring}. In: 2021 IEEE International Conference on Cyber Security and Resilience (CSR). pp. 55--60 (2021). \doi{10.1109/CSR51186.2021.9527927}

\bibitem{8954414}
Gong, L., Cheng, Q.: {Exploiting Edge Features for Graph Neural Networks}. In: 2019 IEEE/CVF Conference on Computer Vision and Pattern Recognition (CVPR). pp. 9203--9211. IEEE Computer Society, Los Alamitos, CA, USA (june 2019). \doi{10.1109/CVPR.2019.00943}, \url{https://doi.ieeecomputersociety.org/10.1109/CVPR.2019.00943}

\bibitem{9001867}
Husain, A., Salem, A., Jim, C., Dimitoglou, G.: {Development of an Efficient Network Intrusion Detection Model Using Extreme Gradient Boosting (XGBoost) on the UNSW-NB15 Dataset}. In: IEEE International Symposium on Signal Processing and Information Technology (ISSPIT). pp.~1--7 (Dec 2019). \doi{10.1109/ISSPIT47144.2019.9001867}

\bibitem{9612213}
Laghari, S.U.A., Manickam, S., Al-Ani, A.K., Rehman, S.U., Karuppayah, S.: {SECS/GEMsec: A Mechanism for Detection and Prevention of Cyber-Attacks on SECS/GEM Communications in Industry 4.0 Landscape}. IEEE Access  \textbf{9},  154380--154394 (Nov 2021). \doi{10.1109/ACCESS.2021.3127515}

\bibitem{9663537}
Li, W., Meng, W., Kwok, L.F.: {Surveying Trust-Based Collaborative Intrusion Detection: State-of-the-Art, Challenges and Future Directions}. IEEE Communications Surveys \& Tutorials  \textbf{24}(1),  280--305 (Dec, 2022). \doi{10.1109/COMST.2021.3139052}

\bibitem{DBLP:journals/corr/abs-2103-16329}
Lo, W.W., Layeghy, S., Sarhan, M., Gallagher, M., Portmann, M.: {E-GraphSAGE: {A} Graph Neural Network based Intrusion Detection System}. CoRR  \textbf{abs/2103.16329} (Apr 2021), \url{https://arxiv.org/abs/2103.16329}

\bibitem{CICIDS2017}
Panigrahi, R., Borah, S.: {A detailed analysis of CICIDS2017 dataset for designing Intrusion Detection Systems}. International Journal of Engineering \& Technology  \textbf{7},  479--482 (Jan 2018)

\bibitem{9564233}
Pei, Y., Huang, T., van Ipenburg, W., Pechenizkiy, M.: {ResGCN: Attention-based Deep Residual Modeling for Anomaly Detection on Attributed Networks}. In: 2021 IEEE 8th International Conference on Data Science and Advanced Analytics (DSAA). pp.~1--2 (2021). \doi{10.1109/DSAA53316.2021.9564233}

\bibitem{https://doi.org/10.48550/arxiv.2202.11097}
{Petar Veličković}: {Message passing all the way up} (Feb 2022). \doi{10.48550/ARXIV.2202.11097}, \url{https://arxiv.org/abs/2202.11097}

\bibitem{DBLP:journals/corr/abs-2107-14756}
Pujol{-}Perich, D., Su{\'{a}}rez{-}Varela, J., Cabellos{-}Aparicio, A., Barlet{-}Ros, P.: {Unveiling the potential of Graph Neural Networks for robust Intrusion Detection}. CoRR  \textbf{abs/2107.14756} (Aug 2021), \url{https://arxiv.org/abs/2107.14756}

\bibitem{9634972}
Qin, K., Zhou, Y., Tian, B., Wang, R.: {AttentionAE: Autoencoder for Anomaly Detection in Attributed Networks}. In: 2021 International Conference on Networking and Network Applications (NaNA). pp. 480--484 (2021). \doi{10.1109/NaNA53684.2021.00089}

\bibitem{9457024}
Shettar, P., Kachavimath, A.V., Mulla, M.M., G, N.D., Hanchinmani, G.: {Intrusion Detection System using MLP and Chaotic Neural Networks}. In: 2021 International Conference on Computer Communication and Informatics (ICCCI). pp.~1--4 (2021). \doi{10.1109/ICCCI50826.2021.9457024}

\bibitem{9202751}
Zerhoudi, S., Granitzer, M., Garchery, M.: {Improving Intrusion Detection Systems using Zero-Shot Recognition via Graph Embeddings}. In: 2020 IEEE 44th Annual Computers, Software, and Applications Conference (COMPSAC). pp. 790--797 (July 2020). \doi{10.1109/COMPSAC48688.2020.0-165}

\end{thebibliography}
\end{document}